\documentclass[reqno]{amsart}
\usepackage{amsmath,amssymb,amsthm}
\usepackage{graphicx,a4wide}
\usepackage{epstopdf}

\theoremstyle{plain}
\newtheorem{thm}{\bf Theorem}
\newtheorem{lem}[thm]{\bf Lemma}
\newtheorem{defn}{\bf Definition}
\theoremstyle{remark}

\newcommand{\prn}[1]{\left(#1\right)}
\newcommand{\pair}[2]{\left<#1,#2\right>_{SS'}}
\newcommand{\R}{\mathbb{R}}
\newcommand{\E}[1]{\mathbb{E}\left[#1\right]}
\renewcommand{\vec}[1]{\mathbf{#1}}
\newcommand{\mat}[1]{\mathbf{#1}}
\newcommand{\ud}[1]{\, \mathrm{d}#1}

\begin{document}
\parindent0ex
\parskip2ex

%=============================================================================================
\title[Optimal prediction for radiative transfer]
{Optimal prediction for radiative transfer: A new perspective on moment closure}

\author{Martin Frank}
\address{Martin Frank \\
RWTH Aachen University \\ Schinkelstr.~2 \\ D-52062 Aachen \\ Germany}
\email{frank@mathcces.rwth-aachen.de}
\urladdr{http://www.mathcces.rwth-aachen.de/doku.php?id=5people:frank:start}

\author{Benjamin Seibold}
\address{Benjamin Seibold \\
Temple University \\ 1805 N.~Broad Street \\ Philadelphia, PA 19122}
\email{seibold@temple.edu}
\urladdr{http://www.math.temple.edu/\~{}seibold}


\thanks{The authors thank Martin Grothaus helpful suggestions on measures in function spaces. The support by the German Research Foundation and the National Science Foundation is acknowledged. M.~Frank was supported by DFG grant KL 1105/14/2. B.~Seibold was partially supported by NSF grants DMS--0813648 and DMS--1007899.
}

%=============================================================================================
\begin{abstract}
Moment methods are classical approaches that approximate the mesoscopic radiative transfer equation by a system of macroscopic moment equations. An expansion in the angular variables transforms the original equation into a system of infinitely many moments. The truncation of this infinite system is the moment closure problem. Many types of closures have been presented in the literature. In this note, we demonstrate that optimal prediction, an approach originally developed to approximate the mean solution of systems of nonlinear ordinary differential equations, can be used to derive moment closures. To that end, the formalism is generalized to systems of partial differential equations. Using Gaussian measures, existing linear closures can be re-derived, such as $P_N$, diffusion, and diffusion correction closures. This provides a new perspective on several approximations done in the process and gives rise to ideas for modifications to existing closures.
\end{abstract}
%=============================================================================================

\maketitle

%=============================================================================================
\section{Introduction}
%=============================================================================================
In many areas, (systems of) macroscopic equations can be derived from mesoscopic kinetic equations. For instance, in the Navier-Stokes and Euler equations, the macroscopic fluid variables, e.g.~density and momentum, are moments of the phase space distribution of the Boltzmann equation. Similarly, in the equation of radiative transfer \cite{Modest1993}, the direction dependent kinetic equations can be transformed into a coupled system of infinitely many moments. Such moment methods can be derived by performing an infinite expansion in the kinetic variable, and then considering only finitely many members of this expansion. While moment methods are by far not the only way to numerically solve the equations of radiative transfer, they play an important role, and a deeper understanding of their problems is important. Recent discussions of the role of moment methods are provided in overview papers by Brunner \cite{Brunner2002}, and by McClarren and Hauck \cite{McClarrenHauck2010}.

Moment methods start with an infinite system of moments that is equivalent to the original equation of radiative transfer. To admit a computation, this infinite system must be approximated by finitely many moments. The challenge to devise approximations that model the influence of the non-considered moments on the considered moments as accurately as possible is the moment closure problem. Commonly known closure strategies are based on truncating and approximating the moment equations, and observing to which extent the solution of the approximate system is close to the true solution. The approximations are supported by physical arguments, such as higher moments being small or adjusting instantaneously. Another strategy is to use asymptotic analysis. In Sect.~\ref{sec:moment_system_radiation}, we outline a commonly used moment hierarchy for radiative transfer, and present some classical linear closure strategies.

In this paper, we propose an alternative strategy: first, an identity for the evolution of the lowest $N$ moments is formulated, and then closures are derived by approximating this identity. Specifically, we show that the method of optimal prediction \cite{ChorinKastKupferman1998_1, ChorinHaldKupferman2000, ChorinHaldKupferman2002, Chorin2003, ChorinHald2006} can be applied to the equation of radiative transfer and yields closed systems of finitely many moments. The idea of optimal prediction, briefly outlined in Sect.~\ref{sec:conditional_expectations}, is to consider the mean solution of a large system, and approximate it by a smaller system that is derived by averaging the equations with respect to an underlying probability measure. The approach can be understood as a way to remove undesired modes, but in an averaged fashion, instead of merely neglecting them. We show that with this formalism, existing linear closures can be re-derived, such as $P_N$, diffusion, and diffusion correction closures.

Optimal prediction is an extension of the Mori-Zwanzig formalism \cite{Mori1965, Zwanzig1980}. It has been applied to partial differential equations \cite{ChorinKastKupferman1999, BellChorinCrutchfield2000}, however, only after reducing them to a system of ordinary differential equations using a Fourier expansion or a semi-discretization step. Here, we generalize the formalism to partial differential equations and measures on function spaces. The idea of using the Mori-Zwanzig formalism for moment closures in radiative transfer was first presented in \cite{SeiboldFrank2009}, however, the derivations done were only formally. In this paper, we show that the ad-hoc derivations done in \cite{SeiboldFrank2009} are in fact justified, by properly defining measures on function spaces (see Sect.~\ref{sec:conditional_expectations}). This generalization is required so that measures on the space of moments can be considered. We restrict ourselves to Gaussian measures. This choice is motivated by the fact that it turns out reproduce existing linear closures, rather than by physical considerations. In Sect.~\ref{sec:linear_optimal_prediction}, we present linear optimal prediction in function spaces. Then, in Sect.~\ref{sec:apply_op_radiation}, we apply linear optimal prediction to the radiation moment system, and show that existing closures can be derived with it. In addition, the approach gives rise to new closures. While the optimal prediction formalism does not remove the arbitrariness in the closure procedure, it introduces it in a rational and comprehensible manner through the choice of a measure and the approximation of an integral. Thus, it provides a new perspective on the errors incurred due to the truncation of the infinite system.

%=============================================================================================
\section{Moment Models for Radiative Transfer}
\label{sec:moment_system_radiation}
%=============================================================================================
The radiative transfer equation (RTE) is \cite{Modest1993}
\begin{equation}
\frac{1}{c}\partial_t I(x,\Omega,t)+\Omega\nabla I(x,\Omega,t)
+\sigma_t(x) I(x,\Omega,t)
=\int_{4\pi}\sigma_s(x,\Omega\cdot\Omega') I(x,\Omega',t)\ud{\Omega'}+q(x,t)\;.
\label{eq:radiative_equation_full}
\end{equation}
In this equation, the radiative intensity $I(x,\Omega,t)$ can be viewed as the the number of particles at time $t$, position $x$, traveling into direction $\Omega$. Equation \eqref{eq:radiative_equation_full} is a mesoscopic phase space equation, modeling loss due to scattering and absorption ($\sigma_t$-term), gain due to anisotropic scattering ($\sigma_s$-term) and containing an emission term $q$. Due to the large number of unknowns, a direct numerical simulation of \eqref{eq:radiative_equation_full} is very costly. Often times only the lowest moments of the intensity with respect to the direction $\Omega$ are of interest. Moment models attempt to approximate \eqref{eq:radiative_equation_full} by a coupled system of moments.

For the sake of notational simplicity, we consider a slab geometry. However, all methods presented here can be generalized. Consider a plate that is finite along the $x$-axis and infinite in the $y$ and $z$ directions. The system is assumed to be invariant under translations in $y$ and $z$ and under rotations around the $x$-axis. In this case the radiative intensity can only depend on the scalar variable $x$ and on the azimuthal angle $\theta = \arccos(\mu)$ between the $x$-axis and the direction of flight. Furthermore, we select units such that $c=1$. The system becomes
\begin{equation}
\partial_t I(x,\mu,t) + \mu \partial_x I(x,\mu,t)  +\sigma_t I(x,\mu,t)
=\int_{-1}^1\sigma_s(x,\mu,\mu') I(x,\mu',t)\ud{\mu'}+q(x,t)\;,
\label{eq:radiation_equation_simple}
\end{equation}
with $t>0$, $x\in\R$, and $\mu\in[-1,1]$. It is supplied with initial conditions $I(x,\mu,0) = \mathring{I}(x,\mu)$. Here, the scattering kernel is
\begin{equation*}
\sigma_s(x,\mu,\mu') = \int_0^{2\pi} \sigma_s(x,\mu\mu'+ \sqrt{1-\mu^2}\sqrt{1-\mu'^2}\cos\varphi) \ud{\varphi}\;.
\end{equation*}
Moment methods start with an infinite system of moments, that is equivalent to the original equation of radiative transfer. Macroscopic approximations to \eqref{eq:radiative_equation_full} can be derived using angular basis functions \cite{Reed1972, BrownChangHanebutteRathkopf2000, BrunnerHolloway2005}, such as spherical harmonics. In 1d slab geometry \eqref{eq:radiation_equation_simple}, Legendre polynomials \cite{Davison1958, Chandrasekhar1960} are commonly used, since they form an orthogonal basis on $[-1,1]$. We assume that the scattering kernel can be expanded into Legendre polynomials
\begin{equation*}
\sigma_s(x,\mu,\mu') = \sum_{n=0}^\infty \tfrac{2n+1}{2}\sigma_{sn}(x) P_n(\mu) P_n(\mu')\;,
\end{equation*}
and introduce the transport coefficients $\sigma_{an} = \sigma_t - \sigma_{sn}$.

To derive a moment system, we define the moments
\begin{equation*}
u_l(x,t) = \int_{-1}^1 I(x,\mu,t)P_l(\mu)\ud{\mu}\;.
\end{equation*}
Multiplying (\ref{eq:radiation_equation_simple}) with $P_k$ and integrating over $\mu$ gives
\begin{equation*}
\partial_t u_k(x,t) + \partial_x \int_{-1}^1 \mu P_k(\mu) I(x,\mu,t)\ud{\mu}+\sigma_t(x) u_k(x,t)
= \sigma_{sk}(x) u_k(x,t) + 2q(x,t)\delta_{k,0}\;.
\end{equation*}
Using the recursion relation for the Legendre polynomials leads to
\begin{equation*}
\partial_t u_k+\partial_x \sum_{l=0}^\infty b_{kl} u_l
= (\sigma_{sk}(x)-\sigma_t(x) ) u_k(x,t) + 2q(x,t)\delta_{k,0}\;.
\end{equation*}
where $b_{kl} = \tfrac{k+1}{2k+1}\delta_{k+1,l}+\tfrac{k}{2k+1}\delta_{k-1,l}$. This is an infinite tridiagonal system for $k=0,1,2,\dots$
\begin{equation}
\partial_t u_k+b_{k,k-1}\partial_x u_{k-1}+b_{k,k+1}\partial_x u_{k+1} = -c_k u_k+q_k
\label{eq:radiative_moment_system_component}
\end{equation}
of first-order partial differential equations. Using the (infinite) matrix notation
\begin{equation*}
\mat{B} = \begin{pmatrix}
0 & 1 & & & \\
\frac{1}{3} & 0 & \frac{2}{3} & & \\
& \frac{2}{5} & 0 & \frac{3}{5} & \\
& & \frac{3}{7} & 0 & \ddots \\ & & & \ddots & \ddots
\end{pmatrix}
\quad\text{,}\quad
\mat{C} = \begin{pmatrix}
\sigma_{a0} & & & \\
& \sigma_{a1} & & \\
& & \sigma_{a2} & \\
& &  & \ddots
\end{pmatrix}
\quad\text{, and}\quad
\mat{q} = \begin{pmatrix} 2 q \\ 0 \\ \vdots \\ \vdots \end{pmatrix}
\label{eq:radiative_moment_matrices}
\end{equation*}
we can write \eqref{eq:radiative_moment_system_component} as
\begin{equation}
\partial_t\vec{u} = \underbrace{-\mat{B}\cdot\partial_x\vec{u}
-\mat{C}\cdot\vec{u}}_{= R\vec{u}}+\vec{q}\;.
\label{eq:radiative_moment_system}
\end{equation}
The infinite moment system \eqref{eq:radiative_moment_system} is equivalent to the transfer equation \eqref{eq:radiation_equation_simple}, since we represent an $L^2$ function in terms of its Fourier components. In order to admit a numerical computation, \eqref{eq:radiative_moment_system} has to be approximated by a system of finitely many moments $I_0,\dots,I_N$, i.e.~all modes $I_l$ with $l>N$ are not considered. In order to obtain a closed system, in the equation for $I_N$, the dependence on $I_{N+1}$ has to be eliminated. A question of fundamental interest is how to close the moment system, i.e.~by what to replace the dependence on $I_{N+1}$. In the following, we provide a brief overview over commonly used closure strategies, and ways to derive and justify them.

%---------------------------------------------------------------------------------------------
\subsection{$P_N$ closure}
\label{subsec:closure_PN}
%---------------------------------------------------------------------------------------------
The simplest closure, the so-called \emph{$P_N$ closure} \cite{Chandrasekhar1944, Reed1972} is based on truncating the sequence $I_l$, i.e.~$I_l=0$ for $l>N$. The physical argument is that if the system is close to equilibrium, then the underlying particle distribution is uniquely determined by the lowest-order moments. For example, the $P_1$ equations are
\begin{align*}
\partial_t u_0 + \partial_x u_1 &= -\sigma_{a0} u_0+ 2q \\
\partial_t u_1 + \tfrac{1}{3}\partial_x u_0 &= - \sigma_{a1} u_1\;.
\end{align*}
The $P_N$ equations can also be derived by asymptotic analysis in the following way \cite{LarsenPomraning1991}. Introduce the dimensionless parameter $\varepsilon$ as the ratio of the total scattering mean free path and a characteristic macroscopic length scale. Furthermore, assume that the coefficients in the RTE asymptotically depend on $\varepsilon$ as
\begin{equation*}
\partial_t I(x,\mu,t) + \mu \partial_x I(x,\mu,t)+\frac{\sigma_t}{\varepsilon} I(x,\mu,t)
=\int_{-1}^1\sigma_s(x,\mu,\mu') I(x,\mu',t)\ud{\mu'}+q(x,t)\;.
\end{equation*}
All other quantities are $\mathcal{O}(1)$. If in addition, the scattering kernel scales as
\begin{equation*}
\sigma_s(x,\mu,\mu') = \sum_{n=0}^N \tfrac{2n+1}{2}\prn{\frac{\sigma_t(x)}{\varepsilon} - \sigma_{an}(x)} P_n(\mu) P_n(\mu') + \sum_{n=N+1}^\infty \tfrac{2n+1}{2}
\frac{\sigma_{sn}(x)}{\varepsilon} P_n(\mu) P_n(\mu')\;,
\end{equation*}
then the $P_N$ equations are an asymptotic limit of the RTE. This means that the solution of the scaled RTE converges to the solution of the $P_N$ equations as $\varepsilon$ tends to zero.

%---------------------------------------------------------------------------------------------
\subsection{Diffusion closure}
\label{subsec:closure_diffusion}
%---------------------------------------------------------------------------------------------
The classical diffusion closure is defined for $N=1$. We assume $I_1$ to be quasi-stationary and neglect $I_l$ for $l>1$, thus the equations read
\begin{align*}
\partial_t u_0 + \partial_x u_1 &= -\sigma_{a0} u_0+2q \\
\tfrac{1}{3}\partial_x u_0 &= - \sigma_{a1} u_1\;.
\end{align*}
Solving the second equation for $u_1$ and inserting it into the first equation yields the diffusion approximation
\begin{equation}
\partial_t u_0 - \partial_x \tfrac{1}{3\sigma_{a1}}\partial_x u_0 = -\sigma_{a0} u_0+2q\;.
\label{eq:closure_diffusion}
\end{equation}
Again, an asymptotic scaling can be used to justify the diffusion approximation \cite{LarsenKeller1974}. The diffusion-scaled RTE is
\begin{equation*}
\varepsilon\partial_t I(x,\mu,t) + \mu \partial_x I(x,\mu,t)  +\frac{\sigma_t}{\varepsilon} I(x,\mu,t)
=\int_{-1}^1\sigma_s(x,\mu,\mu') I(x,\mu',t)\ud{\mu'}+\varepsilon q(x,t)\;,
\end{equation*}
where now the scattering kernel scales as
\begin{equation*}
\sigma_s(x,\mu,\mu') = \tfrac{1}{2}\prn{\frac{\sigma_t(x)}{\varepsilon} - \varepsilon\sigma_{a0}(x)} + \sum_{n=1}^\infty \tfrac{2n+1}{2}
\frac{\sigma_{sn}(x)}{\varepsilon} P_n(\mu) P_n(\mu')\;.
\end{equation*}

%---------------------------------------------------------------------------------------------
\subsection{$D_N$ diffusion correction closure}
%---------------------------------------------------------------------------------------------
A new hierarchy of $P_N$ approximations, denoted \emph{diffusion correction} or $D_N$ closure, has been proposed by Levermore \cite{Levermore2005, SchaeferFrankLevermore2009}. In slab geometry, it can be derived in the following way: We assume that $u_l=0$ for $l>N+1$. Contrary to $P_N$, the $(N+1)$-st moment is assumed to be quasi-stationary. Setting $\partial_t u_{N+1} = 0$ yields the algebraic relation
\begin{equation*}
u_{N+1} = -\tfrac{1}{\sigma_{a,N+1}}\tfrac{N+1}{2N+3}\partial_x u_N\;,
\end{equation*}
which, substituted into the equation for $u_N$, yields an additional diffusion term for the last moment:
\begin{equation}
\partial_t u_N+b_{N,N-1}\partial_x u_{N-1}-\theta\partial_{xx} u_N
= \begin{cases} 2q-\sigma_{a0} u_0 & \text{if~} N=0 \\
-\sigma_{aN} u_N & \text{if~} N>0 \end{cases}
\label{eq:closure_diffusion_correction}
\end{equation}
where $\theta = b_{N,N+1}\frac{1}{\sigma_{a,N+1}}\frac{N+1}{2N+3} = \frac{1}{\sigma_{a,N+1}}\frac{(N+1)^2}{(2N+1)(2N+3)}$. For $N=0$ this closure becomes the classical diffusion closure \eqref{eq:closure_diffusion}. The original derivation \cite{Levermore2005, SchaeferFrankLevermore2009} uses asymptotic scaling arguments similar to the ones above. The $D_1$ equations are
\begin{align*}
\partial_t u_0 + \partial_x u_1 &= -\sigma_{a0} u_0+ 2q \\
\partial_t u_1 + \tfrac{1}{3}\partial_x u_0 &= - \sigma_{a1} u_1 + \tfrac{4}{15\sigma_{a2}}\partial_{xx} u_1\;.
\end{align*}

%---------------------------------------------------------------------------------------------
\subsection{Other types of closures}
%---------------------------------------------------------------------------------------------
Other higher order diffusion approximations exist, such as the so-called \emph{simplified $P_N$} ($SP_N$) equations. These have been derived in an ad hoc fashion \cite{Gelbard1960, Gelbard1961, Gelbard1962} and have subsequently been substantiated via asymptotic analysis \cite{LarsenMorelMcGhee1996} and via a variational approach \cite{TomasevicLarsen1996, BrantleyLarsen2000}. See also the recent review \cite{McClarren2010} and the brief derivation outlined in \cite{FrankKlarPinnau2010}. It is demonstrated in \cite{SeiboldFrank2009} that the formalism presented in this paper can be used to derive certain versions or variations of these types of models.

Many nonlinear approximations exist in the literature, most prominently flux-limited diffusion \cite{Kershaw1976}, variable Eddington factors \cite{Levermore1984, KnollRiderOlson1999, Su2001}, and minimum entropy methods \cite{MullerRuggeri1993, AnilePennisiSammartino1991, DubrocaFeugeas1999, DubrocaFrankKlarThommes2003, FrankKlarLarsenYasuda2007, TurpaultFrankDubrocaKlar2004, FrankDubrocaKlar2006}. Furthermore, approaches based on partial moments exist \cite{TurpaultFrankDubrocaKlar2004, FrankDubrocaKlar2006}. None of these nonlinear closures is considered here.

%=============================================================================================
\section{Conditional Expectations}
\label{sec:conditional_expectations}
%=============================================================================================
Optimal prediction, as introduced by Chorin, Hald, Kast, Kupferman, et~al. \cite{ChorinKastKupferman1998_1, ChorinHaldKupferman2000, ChorinHaldKupferman2002, Chorin2003, ChorinHald2006} seeks the mean solution of a time-dependent system, when only part of the initial data is known, but a measure on the phase space is available. The approach has been developed for ordinary differential equations and applied in various problems that involve dynamical systems \cite{Seibold2004, ChorinStinis2005}. The key ideas of the approach are as follows. Consider a system of ordinary differential equations
\begin{equation}
\vec{\dot x}(t) = \vec{R}(\vec{x}(t)) \ , \quad \vec{x}(0) = \vec{\mathring{x}}\;.
\label{eq:ode_system}
\end{equation}
Let the vector of unknowns be split $\vec{x} = (\vec{x}_{C},\vec{x}_F)$ into the resolved variables $\vec{x}_{C}$ that are of interest, and the unresolved variables $\vec{x}_F$ that should be ``averaged out''.\footnote{In this paper, we borrow a subscript notation from the area of multigrid methods, in which $C$ stands for ``coarse'', and $F$ stands for ``fine''. In the following sections, this notation is also used for block matrices and block operators.} Assume that the initial conditions $\mathring{\vec{x}}_{C}$ for the resolved variables are known, while the initial conditions $\mathring{\vec{x}}_F$ for the unresolved variables are not known or not of relevance. In addition, let a probability density $f(\vec{x})$ be given.

Given the known initial conditions $\mathring{\vec{x}}_{C}$, the measure $f$ induces a conditioned measure $f_{\vec{\mathring{x}}_{C}}(\vec{x}_F) = \tilde Z^{-1} f(\mathring{\vec{x}}_{C},\vec{x}_F)$ for the remaining unknowns, where $\tilde Z = \int f(\mathring{\vec{x}}_{C},\vec{x}_F)\ud{\vec{x}_F}$. An average of a function $u(\vec{x}_{C},\vec{x}_F)$ with respect to $f_{\vec{\mathring{x}}_{C}}$ is the conditional expectation
\begin{equation}
Pu = \E{u|\vec{x}_{C}} = \frac
{\int u(\vec{x}_{C},\vec{x}_F)f(\vec{x}_C,\vec{x}_F)\ud{\vec{x}_F}}
{\int f(\vec{x}_{C},\vec{x}_F)\ud{\vec{x}_F}}\;.
\label{eq:cond_expect_operator}
\end{equation}
It is an orthogonal projection with respect to the inner product $(u,v) = \E{uv}$, which is defined by the expectation $\E{u} = \int\int u(\vec{x}_{C},\vec{x}_F) f(\vec{x}_{C},\vec{x}_F)\ud{\vec{x}_{C}}\ud{\vec{x}_F}$. Let $\varphi(\vec{x},t)$ denote the solution of \eqref{eq:ode_system}, for the initial conditions $\vec{x}=(\vec{x}_{C},\vec{x}_F)$. Then optimal prediction seeks for the mean solution
\begin{equation}
P\varphi(\vec{x},t) = \E{\varphi(\vec{x}_{C},\vec{x}_F,t)|\vec{x}_{C}}\;.
\label{eq:op_mean_solution}
\end{equation}
Optimal prediction provides strategies to formulate systems for $\vec{x}_{C}$ that are faster to compute than solving the original system \eqref{eq:ode_system}. One approach, named first order optimal prediction, is based on applying the conditional expectation \eqref{eq:cond_expect_operator} to the original equation's \eqref{eq:ode_system} right hand side. A related approach is based on the Mori-Zwanzig formalism \cite{Mori1965, Zwanzig1980} in a version for conditional expectations \cite{ChorinHaldKupferman2000}. Applying the formalism to the Liouville equation for \eqref{eq:ode_system} approximates the mean solution by an integro-differential equation, that involves the first order right hand side, plus a memory kernel.

In this paper, we formulate optimal prediction for systems of partial differential equations. In particular, in Sect.~\ref{sec:linear_optimal_prediction}, we apply the Mori-Zwanzig formalism to equations of the type of the radiative moment system \eqref{eq:radiative_moment_system}. In order to extend the projection \eqref{eq:cond_expect_operator} to these types of systems, we have to construct a measure on a suitable infinite-dimensional function space, and an expression for its conditional expectation. Both can usually be achieved by considering a suitable sequence of finite-dimensional measures \cite{HidaKuoPotthoffStreit1993, BerezanskyKondratiev1995}. In addition, a measure can be directly defined by its characteristic functional. Formally, the characteristic functional is given by the measure's Fourier transform. Here we focus on Gaussian measures, because they are one of the few classes where the construction above is possible, and where explicit formulas can be derived. In the following, we present the formalism for conditional expectations first in finitely many dimensions, and then in function spaces.

%---------------------------------------------------------------------------------------------
\subsection{Gaussians in Finite Dimensions}
%---------------------------------------------------------------------------------------------
Due to Bochner's theorem \cite{BerezanskyKondratiev1995}, a measure on $\R^n$ is uniquely determined by its Fourier transform, or characteristic functional
\begin{equation}
\label{eq:Bochner}
\theta(\vec{y}) = \int_{\R^n} f(\vec{x}) \exp(i \vec{y}^T\vec{x}) \ud{\vec{x}}\;.
\end{equation}
Indeed, if
\begin{enumerate}
\item $\theta(\vec{0}) = 1$,
\item $\theta$ is continuous on $\R^n$, and
\item $\theta$ is positive definite in the sense that the matrix
$(\theta(\vec{y}_i-\vec{y}_j))_{i,j=1,\dots,N}$ is positive semi-definite for
all $N$ and all $\vec{y}_i\in\R^n$,
\end{enumerate}
then there is unique probability measure $\lambda$ with density $f$ on the $\sigma$-algebra of Borel sets of $\R^n$ such that (\ref{eq:Bochner}) holds.

Let $\mat{A}\in\R^{n\times n}$ be a symmetric positive definite matrix, and $\vec{m}\in\R^n$. Then
\begin{equation}
f(\vec{x}) = \det(\mat{2\pi A})^{-\frac{1}{2}}
\exp\prn{-\tfrac{1}{2}(\vec{x}-\vec{m})^T \mat{A}^{-1}(\vec{x}-\vec{m})}
\label{eq:Gaussian_measure_Rn}
\end{equation}
is a probability density on $\R^n$. Introducing the inner product generated by $\mat{A}$ as
\begin{equation*}
\langle \vec{x},\vec{y}\rangle_A := \vec{x}^T \mat{A}\vec{y}\;,
\end{equation*}
the Gaussian measure with density (\ref{eq:Gaussian_measure_Rn}) has the characteristic functional
\begin{equation*}
\theta(\vec{y}) = \exp
\prn{-\tfrac{1}{2}\langle\vec{y},\vec{y}\rangle_A+ i\vec{y}^T\vec{m}}\;.
\end{equation*}
This functional satisfies the three conditions above.

The conditional expectation of the Gaussian, given parts of the vector $\vec{x}$, is given by
\begin{lem}
\label{lem:conditional_expectation_finite}
Decompose the vectors $\vec{x}$, $\vec{m}$ and the matrix $\mat{A}$ into
\begin{equation*}
\vec{x} = \begin{bmatrix} \vec{x}_{C} \\ \vec{x}_F \end{bmatrix}
\quad\text{,}\quad
\vec{m} = \begin{bmatrix} \vec{m}_{C} \\ \vec{m}_F \end{bmatrix}
\quad\text{and}\quad
\mat{A} = \begin{bmatrix} \mat{A}_{CC} & \mat{A}_{CF} \\
\mat{A}_{FC} & \mat{A}_{FF} \end{bmatrix}\;.
\end{equation*}
Then the conditional expectation is
\begin{equation}
\E{\vec{x}|\vec{x}_C}
= \begin{bmatrix}\E{\vec{x}_C|\vec{x}_C} \\
\E{\vec{x}_F|\vec{x}_C} \end{bmatrix}
= \begin{bmatrix}
\frac{\int \vec{x}_C f(\vec{x}_C,\vec{x}_F)\ud{\vec{x}_F}}
{\int f(\vec{x}_C,\vec{x}_F)\ud{\vec{x}_F}} \\
\frac{\int \vec{x}_F f(\vec{x}_C,\vec{x}_F)\ud{\vec{x}_F}}
{\int f(\vec{x}_C,\vec{x}_F)\ud{\vec{x}_F}}
\end{bmatrix}
= \begin{bmatrix}
\vec{x}_C \\
\vec{m}_F+\mat{A}_{FC}\mat{A}_{CC}^{-1}
(\vec{x}_C-\vec{m}_C)
\end{bmatrix}\;.
\label{eq:conditional_expectation_finite}
\end{equation}
\end{lem}
\begin{proof}
The identity $\E{\vec{x}_C|\vec{x}_C} = \vec{x}_C$ is trivial. To calculate $\E{\vec{x}_F|\vec{x}_C}$, consider $\mat{M} = \mat{A}^{-1}$, with the same block matrix notation for $\mat{M}$ as for $\mat{A}$. One can easily verify that
\begin{equation*}
(\vec{x}-\vec{m})^T \mat{A}^{-1}(\vec{x}-\vec{m})
= \|\vec{x}_F-\vec{b}_F\|_{\mat{M}_{FF}}^2
+(\vec{x}_C-\vec{m}_C)^T
\prn{\mat{M}_{CC}-\mat{M}_{CF}\mat{M}_{FF}^{-1}\mat{M}_{FC}}
(\vec{x}_C-\vec{m}_C)\;,
\end{equation*}
where $\vec{b}_F = \vec{m}_F-\mat{M}_{FF}^{-1}\mat{M}_{FC}(\vec{x}_C-\vec{m}_C)$, and the norm is defined as $\|\vec{x}_F\|_{\mat{M}_{FF}}^2 = \vec{x}_F^T{\mat{M}_{FF}}\vec{x}_F$. This yields for the conditional expectation
\begin{equation*}
\E{\vec{x}_F|\vec{x}_C}
= \frac{\int\prn{\vec{x}_F-\vec{b}_F+\vec{b}_F}
\exp\prn{-\frac{1}{2}\|\vec{x}_F-\vec{b}_F\|_{\mat{M}_{FF}}^2}
\ud{\vec{x}_F}}
{\int
\exp\prn{-\frac{1}{2}\|\vec{x}_F-\vec{b}_F\|_{\mat{M}_{FF}}^2}
\ud{\vec{x}_F}}
= \vec{b}_F\;.
\end{equation*}
Since $\mat{M} = \mat{A}^{-1}$, the block matrices satisfy $\mat{M}_{FC}\mat{A}_{CC}+\mat{M}_{FF}\mat{A}_{FC} = \mat{0}$, which implies $-\mat{M}_{FF}^{-1}\mat{M}_{FC} = \mat{A}_{FC}\mat{A}_{CC}^{-1}$, and thus proves the claim.
\end{proof}
Expression \eqref{eq:conditional_expectation_finite} coincides with the one given in \cite{ChorinKastKupferman1998_2}. The conditional expectation is a projection, that can be written in the form
\begin{equation*}
P\vec{x} = \E{\vec{x}|\vec{x}_C} = \mat{E}\vec{x}+\mat{F}\vec{m}\;,
\end{equation*}
using the projection matrices
\begin{equation}
\mat{E} = \begin{bmatrix} \mat{I} & \mat{0} \\
\mat{A}_{FC}\mat{A}_{CC}^{-1} & \mat{0} \end{bmatrix}
\quad\text{and}\quad
\mat{F} = \begin{bmatrix} \mat{0} & \mat{0} \\
-\mat{A}_{FC}\mat{A}_{CC}^{-1} & \mat{I} \end{bmatrix}\;.
\label{eq:projection_matrices}
\end{equation}
The orthogonal complement is then
\begin{equation*}
Q\vec{x} = (I-P)\vec{x} = \mat{F}\vec{x}-\mat{F}\vec{m}\;.
\end{equation*}
One can easily verify that $P^2 = P$, $Q^2 = Q$, and $PQ = QP = 0$. If the measure is centered around the origin, i.e.~$\vec{m} = 0$, the projections become simple matrix multiplications $P = \mat{E}$ and $Q = \mat{F}$.

%---------------------------------------------------------------------------------------------
\subsection{Gaussians in Function Spaces}
%---------------------------------------------------------------------------------------------
The construction of measures on spaces of functions uses the characteristic functional. Formally, all expressions from the finite-dimensional case generalize to the infinite-dimensional case. There are some mathematical subtleties related to this construction. For the interested reader, we collect these in this section.

Following \cite{HidaKuoPotthoffStreit1993, BerezanskyKondratiev1995}, we construct measures on the dual $S'$ of a Hilbert space $S$ of functions. This construction is based on the Bochner-Minlos theorem. Its key assumption is that $S$ is nuclear, i.e.\ the identity in $S$ is of Hilbert-Schmidt type. In that case, the three conditions from above on a characteristic functional are necessary and sufficient for the existence of a corresponding measure. The proof uses a sequence of finite-dimensional measures and Bochner's theorem. The following construction of a nuclear Hilbert space serves our latter purposes.
\begin{defn}
Let $\mat{A}$ be a symmetric positive definite infinite matrix (i.e.\ all finite submatrices are symmetric positive definite) such that
\begin{equation*}
\sum_{i,j=1}^\infty |\mat{A}_{ij}|^2 < \infty\;,
\end{equation*}
and let $V$ be a Hilbert space. We define the Hilbert space $l^2_A(V)$, consisting of an infinite vector of elements of $V$, by the inner product
\begin{equation*}
\langle \vec{f},\vec{g}\rangle_{l^2_A(V)}
:= \sum_{i,j=1}^\infty \langle \mat{A}_{ij}\vec{f}_j,\vec{g}_i \rangle_V\;.
\end{equation*}
\end{defn}
In our case, we consider $X=l^2_A(L^2(\R))$. In order to obtain a Gelfand triple $S\subset X\subset S'$ with the nuclearity property, we have to construct a space of smooth test functions and its dual space of distributions. There are several ways to do this. The following construction is standard and frequently used \cite{BerezanskyKondratiev1995}.
\begin{defn}
Let
\begin{equation*}
H = -\frac{d^2}{dx^2} + x^2 + 1\;,
\end{equation*}
and define the Hilbert space $\mathcal{H}(\R)$ as the completion of $C_0^\infty(\R)$ in $L^2(\R)$ with respect to the inner product
\begin{equation*}
\langle f,g \rangle_{\mathcal{H}(\R)} := \langle Hf,g \rangle_{L^2(\R)}\;.
\end{equation*}
\end{defn}
Let $\mathcal{H}_{-1}(\R)$ denote the dual of $\mathcal{H}(\R)$. Then with $S= l^2_{A}(\mathcal{H}(\R))$ and $S'= l^2_{A}(\mathcal{H}_{-1}(\R))$, we have a Hilbert-Schmidt embedding $S\subset X \subset S'$. Thus by the Bochner-Minlos theorem \cite{BerezanskyKondratiev1995}, the characteristic functional
\begin{equation*}
\theta(\vec{f}) = \int_{S'} \exp(i\pair{\vec{f}}{\vec{g}}) \ud{\lambda(\vec{g})}
= \exp\prn{-\tfrac{1}{2}\langle \vec{f},\vec{f} \rangle_X}
\end{equation*}
defines a unique probability measure on $S'$ which satisfies
\begin{equation*}
\int_{S'} \exp(i\pair{\vec{f}}{\vec{g}}) \ud{\lambda(\vec{g})} = \theta(\vec{f})\;.
\end{equation*}
In addition, a nonzero expectation value can be taken into account by noting that the modified functional can be written in terms of an integral which easily allows checking the properties in the Bochner-Minlos theorem:
\begin{align*}
\int_{S'} \exp(i\pair{\vec{f}}{\vec{g}+\vec{m}}) \ud{\lambda(\vec{g})}
&= \int_{S'} \exp(i\pair{\vec{f}}{\vec{g}}) \ud{\lambda(\vec{g})}
\exp(i\pair{\vec{f}}{\vec{m}}) \\
&= \exp\prn{-\tfrac{1}{2}\langle \vec{f},\vec{f} \rangle_X +i\pair{\vec{f}}{\vec{m}}}\;.
\end{align*}
Thus we have
\begin{lem}
Given $\vec{m}\in S'$, the characteristic functional
\begin{equation}
\theta(\vec{f}) = \exp\prn{-\tfrac{1}{2}\langle \vec{f},\vec{f} \rangle_X +i\pair{\vec{f}}{\vec{m}}}
\label{eq:Gaussian_measure_function}
\end{equation}
defines a unique probability measure on $S'$ such that
\begin{equation*}
\int_{S'} \exp(i\pair{\vec{f}}{\vec{g}}) \ud{\lambda(\vec{g})} = \theta(\vec{f})\;.
\end{equation*}
\end{lem}
In the same way as for the measure, certain moments or conditional expectations can be inherited from the finite-dimensional case.
\begin{lem}
Decompose the vector-valued distribution $\vec{u}\in S'$, the vector-valued expectation value $\vec{m}\in S'$ and the matrix $\mat{A}$ into
\begin{equation*}
\vec{u} = \begin{bmatrix} \vec{u}_C \\ \vec{u}_F \end{bmatrix}
\quad\text{,}\quad
\vec{m} = \begin{bmatrix} \vec{m}_C \\ \vec{m}_F \end{bmatrix}
\quad\text{and}\quad
\mat{A} = \begin{bmatrix} \mat{A}_{CC} & \mat{A}_{CF} \\
\mat{A}_{FC} & \mat{A}_{FF} \end{bmatrix}\;.
\end{equation*}
We denote by $\lambda(\vec{u}_F)$ the conditioned Gaussian measure with respect to $\vec{u}_F$, given $\vec{u}_C$. Then for all vector-valued test functions $\vec{f} = \begin{bmatrix} \vec{f}_C , \vec{f}_F \end{bmatrix}^T$, we have the conditional expectations
\begin{align*}
\int \pair{\vec{f}_C}{\vec{u}_C} \ud{\lambda(\vec{u}_F)}
&= \pair{\vec{f}_C}{\vec{u}_C} \\
\int \pair{\vec{f}_F}{\vec{u}_F} \ud{\lambda(\vec{u}_F)}
&= \pair{\vec{f}_F}{\vec{m}_F}
+\pair{\vec{f}}{\mat{A}_{FC}\mat{A}_{CC}^{-1}
(\vec{u}_C-\vec{m}_C)}\;.
\end{align*}
Here, $\int \pair{\vec{f}_C}{\vec{u}_C} \ud{\lambda(\vec{u}_F)}$ and $\int \pair{\vec{f}_F}{\vec{u}_F} \ud{\lambda(\vec{u}_F)}$ are the weak formulations of the integrals $\int \vec{u}_C \ud{\lambda(\vec{u}_F)}$ and $\int \vec{u}_F \ud{\lambda(\vec{u}_F)}$, which can be interpreted as conditional expectations of an $S'$-valued random variable with probability distribution $\lambda(\vec{u}_F)$.
\end{lem}
\begin{proof}
The proof follows the proof of the Bochner-Minlos theorem. We approximate the infinite-dimensional Gaussian measure by a sequence of finite-dimensional Gaussian measures. For each of these, we have a formula for the conditional expectation, given by Lemma~\ref{lem:conditional_expectation_finite}. By showing that the limit of conditional measures exists and by showing that the monomials are measurable, we obtain the conditional expectations above.
\end{proof}

As in the finite dimensional case, the projection can be written in matrix form
\begin{equation}
\pair{\vec{f}}{P\vec{u}} =
\int \pair{\vec{f}}{\vec{u}}\ud{\lambda(\vec{u}_F)}
= \pair{\vec{f}}{\mat{E}\vec{u}}+\pair{\vec{f}}{\mat{F}\vec{m}}\;,
\label{eq:conditional_expectation_function}
\end{equation}
using the same projection matrices \eqref{eq:projection_matrices} as in the finite dimensional case. As a short notation, we write \eqref{eq:conditional_expectation_function} as $P\vec{u} = \mat{E}\vec{u}+\mat{F}\vec{m}$, or $P = \mat{E}$ in the case $\vec{m}=\vec{0}$. In the following, whenever we use this short notation, we always mean in the weak sense \eqref{eq:conditional_expectation_function}.

%=============================================================================================
\section{Linear Optimal Prediction}
\label{sec:linear_optimal_prediction}
%=============================================================================================
We now apply optimal prediction to a linear evolution equation under a Gaussian measure. As derived in Sect.~\ref{sec:conditional_expectations}, the conditional expectation is an affine linear projection. Here, we consider a Gaussian centered around the origin, thus $P = \mat{E}$. While this choice is reasonable in many cases, its main purpose is to simplify notation. The results transfer to the case $\vec{m}\neq\vec{0}$, with affine linear transformations instead of matrix multiplications. We present the Mori-Zwanzig formalism \cite{Mori1965, Zwanzig1980} for a linear evolution equation
\begin{equation}
\partial_t\vec{u} = R\vec{u} \ , \quad \vec{u}(0) = \vec{\mathring{u}}\;,
\label{eq:linear_system}
\end{equation}
where $\vec{u}$ is a vector-valued distribution and $R$ is a linear differential operator (or $\vec{u}$ is a vector and $R$ is a matrix, for an ordinary differential equation) that is independent of space and time. Consider a Gaussian measure, defined by a symmetric positive definite matrix $\mat{A}$ (see Sect.~\ref{sec:conditional_expectations}). Let the unknowns and the corresponding operators/matrices be split
\begin{equation*}
\vec{u} = \begin{bmatrix} \vec{u}_C \\ \vec{u}_F \end{bmatrix}
\quad\text{,}\quad
R = \begin{bmatrix} R_{CC} & R_{CF} \\ R_{FC} & R_{FF} \end{bmatrix}
\quad\text{, and}\quad
\mat{A} = \begin{bmatrix}
\mat{A}_{CC} & \mat{A}_{CF} \\ \mat{A}_{FC} & \mat{A}_{FF}
\end{bmatrix}\;.
\end{equation*}
The conditional expectation of the coordinate vector $\vec{u}$ is
\begin{equation*}
P\vec{u} = \E{\vec{u}|\vec{u}_C} = \mat{E}\cdot\vec{u}\;,
\end{equation*}
where we have the projection matrix \eqref{eq:projection_matrices}
\begin{equation*}
\mat{E} = \begin{bmatrix} \mat{I} & \mat{0} \\
\mat{A}_{FC}\mat{A}_{CC}^{-1} & \mat{0} \end{bmatrix}\;.
\end{equation*}
Also, define $\mat{F} = \mat{I}-\mat{E}$ as the orthogonal projection matrix. Due to linearity, for any matrix vector product $\mat{B}\vec{u}$, the projection always applies to the vector itself
\begin{equation}
P\mat{B}\vec{u} = \E{\mat{B}\vec{u}|\vec{u}_C} = \mat{B}\cdot\E{\vec{u}|\vec{u}_C}
= \mat{B}\cdot\mat{E}\cdot\vec{u} = \mathcal{B}\cdot\vec{u}\;,
\label{eq:projected_matrix}
\end{equation}
where the projected matrix takes the form
\begin{equation}
\mathcal{B}
= \begin{bmatrix}
\mat{B}_{CC}+\mat{B}_{CF}\mat{A}_{FC}\mat{A}_{CC}^{-1} & \mat{0} \\
\mat{B}_{FC}+\mat{B}_{FF}\mat{A}_{FC}\mat{A}_{CC}^{-1} & \mat{0}
\end{bmatrix}\;.
\label{eq:op_projected_matrix}
\end{equation}
Let the solution operator of \eqref{eq:linear_system} be denoted by $e^{tR}$. In addition, we consider the solution operator $e^{tR\mat{F}}$ to the orthogonal dynamics equation
\begin{equation}
\partial_t\vec{u} = R\mat{F}\vec{u} \ , \quad \vec{u}(0) = \vec{\mathring{u}}\;.
\label{eq:orthogonal_dynamics}
\end{equation}
We assume that both the original system \eqref{eq:linear_system} and the orthogonal dynamics equation \eqref{eq:orthogonal_dynamics} are well posed. The existence of solutions to the orthogonal dynamics has been proved for $R$ the Liouville operator to a nonlinear differential equation \cite{GivonHaldKupferman2004}. If $R$ is a matrix, then both $e^{tR}$ and $e^{tR\mat{F}}$ are in fact matrix exponentials. For $R$ a differential operator, they stand as a notation for the solution operators generated by $R$ and $R\mat{F}$.
\begin{thm}[Dyson's formula]
Let $R$ be a differential operator and $\mat{E}+\mat{F} = \mat{I}$ a pair of orthogonal projection matrices. Let $e^{tR}$ and $e^{tR\mat{F}}$ denote the evolution operators generated by $R$ respectively $R\mat{F}$. Then
\begin{equation*}
e^{tR} = e^{tR\mat{F}}+\int_0^t e^{(t-s)R\mat{F}}R\mat{E}e^{sR}\ud{s}\;.
\end{equation*}
\end{thm}
\begin{proof}
Define the evolution operator
\begin{equation}
M(t) = e^{tR}-e^{tR\mat{F}}-\int_0^t e^{(t-s)R\mat{F}}R\mat{E}e^{sR}\ud{s}\;.
\label{eq:dyson_formula}
\end{equation}
Its time derivative equals
\begin{equation*}
\partial_t M(t) = Re^{tR}-R\mat{F}e^{tR\mat{F}}-R\mat{E}e^{tR}
-R\mat{F}\int_0^t e^{(t-s)R\mat{F}}R\mat{E}e^{sR}\ud{s} = R\mat{F}M(t)\;. \\
\end{equation*}
With the initial conditions $M(0) = 0$, we obtain that $M(t) = 0 \ \forall t\ge 0$.
\end{proof}
Differentiating \eqref{eq:dyson_formula} with respect to time yields an identity for the solution operator
\begin{equation}
\begin{split}
\partial_t\prn{e^{tR}}
&= R\mat{F}e^{tR\mat{F}}+R\mat{E}e^{tR}
+R\mat{F}\int_0^t e^{(t-s)R\mat{F}}R\mat{E}e^{sR}\ud{s} \\
&= \mathcal{R}e^{tR}+e^{tR\mat{F}}R\mat{F}+\int_0^t K(t-s)e^{sR}\ud{s}\;,
\end{split}
\label{eq:solution_operator_identity}
\end{equation}
where $\mathcal{R} = R\mat{E}$ is the projected differential operator, and $K(t) = e^{tR\mat{F}}R\mat{F}R\mat{E}$ is a memory kernel for the dynamics. Matrix $\mat{E}$ has zeros in the right column, thus $\mathcal{R}$ and $K$ have the same structure
\begin{equation*}
\mathcal{R} =
\begin{bmatrix} \mathcal{R}_{CC} & 0 \\
\mathcal{R}_{FC} & 0 \end{bmatrix}
\quad\text{and}\quad
K = \begin{bmatrix} K_{CC} & 0 \\ K_{FC} & 0 \end{bmatrix}\;.
\end{equation*}
As defined by \eqref{eq:op_mean_solution}, the mean solution of \eqref{eq:linear_system} with respect to the measure defined by \eqref{eq:Gaussian_measure_function} is obtained by applying the projection operator to the solution operator. Since the solution operator is linear, property \eqref{eq:projected_matrix} yields
\begin{equation*}
\vec{u}^\mathrm{m}(t) = Pe^{tR}\vec{\mathring{u}}
= e^{tR}\mat{E}\vec{\mathring{u}}\;,
\end{equation*}
i.e.~the mean solution is a particular solution, obtained by evolving the projected (averaged) initial conditions. Thus, the mean solution operator is $e^{tR}\mat{E}$. Multiplying the identity \eqref{eq:solution_operator_identity} from the right by $\mat{E}$ yields an identity for the mean solution operator
\begin{equation*}
\partial_t\prn{e^{tR}\mat{E}} = \mathcal{R}e^{tR}\mat{E}+
\underbrace{e^{tR\mat{F}}R\mat{F}\mat{E}}_{=0}
+\int_0^t K(t-s)e^{sR}\mat{E}\ud{s}\;,
\end{equation*}
in which the middle term cancels out, since $\mat{F}\mat{E} = \mat{0}$. This yields a new evolution equation for the mean solution
\begin{equation*}
\begin{cases}
\partial_t\vec{u}^\mathrm{m}(t)
&= \mathcal{R}\vec{u}^\mathrm{m}(t)+\int_0^t K(t-s)\vec{u}^\mathrm{m}(s)\ud{s} \\
\vec{u}^\mathrm{m}(0) &= \mat{E}\vec{\mathring{u}}
\end{cases}\;,
\end{equation*}
which reads in block-components
\begin{align}
\partial_t\vec{u}_C^\mathrm{m}
= \mathcal{R}_{CC}\vec{u}_C^\mathrm{m}
+K_{CC}*\vec{u}_C^\mathrm{m}\;, \quad
&\vec{u}_C^\mathrm{m}(0) = \mathring{\vec{u}}_C
\label{eq:op_evolution_interest} \\
\partial_t\vec{u}_F^\mathrm{m}
= \mathcal{R}_{FC}\vec{u}_C^\mathrm{m}
+K_{FC}*\vec{u}_C^\mathrm{m}\;, \quad
&\vec{u}_F^\mathrm{m}(0) =
\mat{A}_{FC}\mat{A}_{CC}^{-1}\mathring{\vec{u}}_C\;.
\label{eq:op_evolution_averaged}
\end{align}
We have derived an equation \eqref{eq:op_evolution_interest} in which the dynamics for the variables of interest is independent of the evolution of the averaged variables. The latter are typically not of interest, but if desired, they can be obtained by integrating \eqref{eq:op_evolution_averaged}. For nonlinear systems of ordinary differential equations, an analogous integro-differential equation can be derived, which approximates the true mean solution. In that context \cite{ChorinHaldKupferman2001}, it is denoted \emph{second order optimal prediction}. For the linear problem considered here, equation \eqref{eq:op_evolution_interest} yields the true mean solution, hence we call it \emph{full optimal prediction}.

The simplest approximation to \eqref{eq:op_evolution_interest}, called \emph{first order optimal prediction}, is obtained by neglecting the convolution term, i.e.~by solving the system
\begin{equation}
\partial_t\vec{u}_C^\mathrm{foop}
= \mathcal{R}_{CC}\vec{u}_C^\mathrm{foop} \ , \quad
\vec{u}_C^\mathrm{foop}(0) = \mathring{\vec{u}}_C\;.
\label{eq:op_first_order}
\end{equation}
A better approximation can be obtained if a time scale $\tau$ exists, after which the kernel becomes negligible: $K(t)\ll K(0)\;\forall t>\tau$. Assuming that $u^{\text{m}} = O(1)$ over the time scale of integration, a piecewise-constant quadrature rule yields the approximation
\begin{equation}
\begin{split}
\int_0^t K(t-s)\vec{u}^\mathrm{m}(s)\ud{s} &= \int_0^t K(s)\vec{u}^\mathrm{m}(t-s)\ud{s}
\approx \int_0^\tau K(s)\vec{u}^\mathrm{m}(t-s)\ud{s} \\
&\approx \int_0^\tau K(0)\vec{u}^\mathrm{m}(t)\ud{s}
= \tau R\mat{F}R\mat{E}\vec{u}^\mathrm{m}(t)\;,
\end{split}
\label{eq:op_second_order_integral}
\end{equation}
which leads to the \emph{second order optimal prediction} system
\begin{equation}
\partial_t \vec{u}_C^\mathrm{soop}
= \mathcal{R}_{CC}\vec{u}_C^\mathrm{soop}
+\tau\prn{R\mat{F}R\mat{E}}_{CC}\vec{u}_C^\mathrm{soop}\ , \quad
\vec{u}_C^\mathrm{soop}(0) = \mathring{\vec{u}}_C\;.
\label{eq:op_second_order}
\end{equation}
Here
\begin{equation*}
\prn{R\mat{F}R\mat{E}}_{CC} =
R_{CF}R_{FC}
+R_{CF}R_{FF}\mat{A}_{FC}\mat{A}_{CC}^{-1}
-R_{CF}\mat{A}_{FC}\mat{A}_{CC}^{-1}R_{CC}
-R_{CF}\mat{A}_{FC}\mat{A}_{CC}^{-1}R_{CF}
\mat{A}_{FC}\mat{A}_{CC}^{-1}
\end{equation*}
is a new linear differential operator, which is second order if $R$ is a first order operator. While this piecewise constant approximation to the memory term is more accurate than not considering the memory term at all, it is not very accurate as a quadrature rule. Its use here is justified for two reasons:
\begin{itemize}
\item First, we wish to restrict our analysis to systems that involve only the current state of the system (i.e.~no delay);
\item Second, this approximation turns out to yield the diffusion-correction closure \eqref{eq:closure_diffusion_correction}, when applied to radiative transfer (see Sect.~\ref{sec:apply_op_radiation}).
\end{itemize}

%=============================================================================================
\section{Application to the Radiation Moment System}
\label{sec:apply_op_radiation}
%=============================================================================================
We now turn our attention to the infinite moment system \eqref{eq:radiative_moment_system}. For consistency with the notation developed in Sect.~\ref{sec:linear_optimal_prediction}, we denote the (infinite) vector of moments by $\vec{u}(x,t) = \prn{u_0(x,t),u_1(x,t),\dots}^T$. In addition, we neglect the source term, since it is unaffected by truncation. The radiation system \eqref{eq:radiative_moment_system} can be written as
\begin{equation*}
\partial_t\vec{u} = R\vec{u}\;,
\label{eq:radiative_moment_system_sourcefree}
\end{equation*}
where the differential operator $R = -\mat{B}\partial_x-\mat{C}$ involves the (infinite) matrices \eqref{eq:radiative_moment_matrices}. As introduced in Sect.~\ref{sec:conditional_expectations}, we consider a Gaussian measure on the space of unknowns, defined by an (infinite) matrix $\mat{A}$. In Sect.~\ref{subsec:apply_op_radiation_foop}, we consider first order optimal prediction, and in Sect.~\ref{subsec:apply_op_radiation_soop}, we consider second order optimal prediction.

%---------------------------------------------------------------------------------------------
\subsection{First Order Optimal Prediction}
\label{subsec:apply_op_radiation_foop}
%---------------------------------------------------------------------------------------------
We wish to truncate the system after the $N$-th component. The system and the measure are split into blocks
\begin{equation*}
\vec{u} = \begin{bmatrix} \vec{u}_C \\ \vec{u}_F \end{bmatrix}
\quad\text{,}\quad
\mat{B} = \begin{bmatrix}
\mat{B}_{CC} & \mat{B}_{CF} \\
\mat{B}_{FC} & \mat{B}_{FF}
\end{bmatrix}
\quad\text{,}\quad
\mat{C} = \begin{bmatrix}
\mat{C}_{CC} & \mat{C}_{CF} \\
\mat{C}_{FC} & \mat{C}_{FF}
\end{bmatrix}
\quad\text{, and}\quad
\mat{A}=\begin{bmatrix} \mat{A}_{CC} & \mat{A}_{CF} \\
\mat{A}_{FC} & \mat{A}_{FF} \end{bmatrix}\;.
\end{equation*}
For the radiation system, we have
\begin{equation*}
\mat{B}_{CF}
= \begin{pmatrix}
0 & \hdots & 0 \\
\vdots & \ddots & \vdots \\
\frac{N+1}{2N+1} & \hdots & 0
\end{pmatrix}
\quad\text{,}\quad
\mat{C}_{CF} = \mat{0}
\quad\text{, and}\quad
\mat{C}_{FC} = \mat{0}\;.
\end{equation*}
Due to \eqref{eq:op_projected_matrix}, the projected differential operator's upper left block is
\begin{equation}
\mathcal{R}_{CC}
= -\prn{\mat{B}_{CC}+\mat{B}_{CF}\mat{A}_{FC}\mat{A}_{CC}^{-1}}\partial_x
-\mat{C}_{CC}\;.
\label{eq:radiation_linear_closure}
\end{equation}
The modification term $\mat{B}_{CF}\mat{A}_{FC}\mat{A}_{CC}^{-1}$ has nonzero entries only in its last row. Hence, first order optimal prediction yields a true closure relation, since only the last equation is modified. The modification is $\frac{N+1}{2N+1}$ times the first row of $\mat{A}_{FC}\mat{A}_{CC}^{-1}$, i.e.~the closure depends solely on the correlations between the moments, given by the measure. We can see that, depending on the choice of $\mat{A}$, first-order optimal prediction can generate all possible linear hyperbolic closures.

The measure enables us to encode a correlation between the resolved moments $\vec{u}_C$ and the averaged moments $\vec{u}_F$, that could come from additional knowledge about the particular setup of the problem. If no such additional knowledge is at hand, it is reasonable to prescribe no correlation between $\vec{u}_C$ and $\vec{u}_F$, by considering a decoupled measure, i.e.~$\mat{A}_{FC} = \mat{0}$. In this case, the system is plainly cut off, and thus the classical $P_N$ closure is obtained.

%---------------------------------------------------------------------------------------------
\subsection{Second Order Optimal Prediction}
\label{subsec:apply_op_radiation_soop}
%---------------------------------------------------------------------------------------------
In Sect.~\ref{subsec:apply_op_radiation_foop}, we have seen that first order optimal prediction with a decoupled Gaussian measure yields the classical $P_N$ closure. For the same measure, we now consider the memory term. Since $\mat{A}_{FC} = \mat{0}$, we have
\begin{equation*}
R\mat{E}
= \begin{bmatrix} R_{CC} & 0 \\ R_{FC} & 0 \end{bmatrix}
\quad\text{,}\quad
R\mat{F}
= \begin{bmatrix} 0 & R_{CF} \\ 0 & R_{FF} \end{bmatrix}
\quad\text{, and}\quad
R\mat{F}R\mat{E} =
\begin{bmatrix} R_{CF}R_{FC} & 0 \\
R_{FF}R_{FC} & 0 \end{bmatrix}\;.
\end{equation*}
For the radiation system \eqref{eq:radiative_moment_system_sourcefree} we get
\begin{equation*}
R\mat{F}R\mat{E}
= \begin{bmatrix} \mat{B}_{CF}\mat{B}_{FC}\partial_{xx} & \mat{0} \\
\mat{B}_{FF}\mat{B}_{FC}\partial_{xx}
+\mat{C}_{FF}\mat{B}_{FC}\partial_{x} & \mat{0} \end{bmatrix}\;,
\end{equation*}
where
\begin{equation*}
\mat{B}_{CF}\mat{B}_{FC}
= \begin{pmatrix}
0 & \hdots & 0 \\
\vdots & \ddots & \vdots \\
0 & \hdots & \frac{(N+1)^2}{(2N+1)(2N+3)}
\end{pmatrix}\;.
\end{equation*}
Formally, the memory term uses solution values at all previous times. However, the different solution components decay at the rates $\sigma_{ai}$. This yields time scales $\tau_i = \frac{1}{\sigma_{ai}}$, over which information from the past can be seen in the solution ($\tau_i$ are a time scales since we have set $c=1$). If we single out one time scale $\tau$, the second order optimal prediction approximation \eqref{eq:op_second_order} becomes
\begin{equation*}
\int_0^t K_{CC}(s)\vec{u}_C(t-s)\ud{s}
\approx\tau\prn{R\mat{F}R\mat{E}}_{CC}\vec{u}_C
= \begin{pmatrix}
0 \\
\vdots \\
0 \\
\tau\frac{(N+1)^2}{(2N+1)(2N+3)}\partial_{xx} u_N
\end{pmatrix}\;.
\end{equation*}
Compared to the $P_N$ closure, a diffusion term is added into the last component of the truncated system. If we identify $\tau$ with $\tau_{N+1}$, i.e.\ the time scale given by the rate of decay of the first truncated moment, then this system is exactly the diffusion correction closure \eqref{eq:closure_diffusion_correction} by Levermore \cite{Levermore2005}, as outlined in Sect.~\ref{subsec:closure_diffusion}. In the case $N=0$, it is equivalent to the classical diffusion approximation.

Although here we have assumed spatially homogeneous coefficients, we expect that the equations can be adapted to the space-dependent case in analogy to diffusion theory. Specifically, if $\sigma_{ai}(x)$ are space dependent, we define $\tau_i(x) = \frac{1}{\sigma_{ai}(x)}$, and replace $\tau_{N+1}\partial_{xx} u_N$ by $\partial_x\prn{\tau_{N+1}(x)\partial_x u_N}$. The validity of this approximation will be addressed in future research.

%=============================================================================================
\section{Discussion}
%=============================================================================================
We have formulated the approach of optimal prediction for a system of linear partial differential equations with an underlying Gaussian measure. An identity for the evolution of a finite number of moments is obtained. Approximations to this identity yield different closures for a truncated version of the full system. The application of the formalism to the equation of radiative transfer allows the re-derivation of classical linear closures, such as $P_N$, diffusion, and diffusion correction. While traditionally these closures are derived using physical arguments or by asymptotic analysis, the optimal prediction formalism generates the closures by choosing a measure and approximating a mathematical identity. This connection yields a new interpretation of diffusion as being connected to the memory of the system (or loss thereof).

If we follow this interpretation, we observe a possible modification of the second order optimal prediction approximation \eqref{eq:op_second_order} (and thus of the diffusion approximation), that may be more accurate for short times: for $t<\tau$, the integral in \eqref{eq:op_second_order_integral} cannot stretch over the whole length $\tau$. Thus, one should replace the coefficient $\tau$ by a time dependent function $f(t)$ that vanishes for $t\to 0$, increases for $0<t<\tau$, and approaches $\tau$ for $t\gg\tau$. For example, a better approximation than \eqref{eq:op_second_order_integral} is given by
\begin{equation}
\int_0^t K(t-s)\vec{u}^\mathrm{m}(s)\ud{s}
\approx\min\{t,\tau\}R\mat{F}R\mat{E}\vec{u}^\mathrm{m}(t)\;,
\label{eq:op_second_order_integral_crescendo}
\end{equation}
which leads to replacing $\tau$ by $\min\{t,\tau\}$ in the second order system \eqref{eq:op_second_order}. While \eqref{eq:op_second_order} yields a classical diffusion correction approximation, expression \eqref{eq:op_second_order_integral_crescendo} leads to a new approximation, which we have called \emph{crescendo-diffusion correction} approximation in a previous paper \cite{SeiboldFrank2009}. The function $\min\{t,\tau\}$ is just one of many possible time dependent coefficients for the second order term. A smoother evolution is derived in \cite{SeiboldFrank2009} by approximating the orthogonal dynamics more accurately. It yields the diffusion function $f(t) = \tau\prn{1-e^{-t/\tau}}$.

The interpretation that diffusion is connected to the memory of the system leads us to a gradual ramp-up of the diffusion coefficient. This suggests that there is an initial layer over which the particle system becomes diffusive. On the one hand, standard techniques do not predict an initial layer. The ad-hoc condition is to just take the moments of the initial condition. This condition also comes out of asymptotic analysis (for diffusion in \cite{LarsenKeller1974}, for $P_N$ in \cite{LarsenPomraning1991}). Computational results \cite{SeiboldFrank2009}, on the other hand, show that there can be a non-negligible initial layer, and that the crescendo diffusion idea improves the quality of the solution. The crescendo modification introduces an explicit time-dependence. The physical rationale is that at $t=0$, the state of the resolved moments is known exactly. Information is lost as time evolves, due to the approximation.

In this paper, optimal prediction with respect to Gaussian measures is considered. However, many other measures are conceivable, for instance a construction in the following spirit. Gaussian measures are supported everywhere, and thus do not respect the conditions on moments to be realizable as moments of a non-negative radiative intensity function \cite{Kershaw1976}. As shown in \cite{KarlinShapley1953, Skibinsky1967}, in the reduced moments $N_k = \frac{u_k}{u_0}\;\forall\,k\ge 1$, this realizability domain is contained in a box of square-summable edge lengths, i.e.~$N_k\in [N_k^-,N_k^+]$ with $N_k^+-N_k^- \le 2^{-2k+2}$. Due to this property it is plausible that measures can be defined (e.g., a uniform distribution in $N_k\;\forall\,k\ge 1$) on the space of infinitely many moments, that vanish outside the realizability domain. Closures derived as conditional expectations with respect to such measures would---in contrast to Gaussian measures---be always realizable and, as a consequence, nonlinear.

Hence, it is an important question whether existing nonlinear closures, such as flux-limited diffusion \cite{Levermore1984} and minimum entropy closures \cite{MullerRuggeri1993, AnilePennisiSammartino1991, DubrocaFeugeas1999} can be derived by optimal prediction with a non-Gaussian measure. If possible, the approach could be used to systematically derive diffusion correction terms for these nonlinear closures. Another fundamental question is to which extent boundary conditions can be included into the formalism. Many other types of closures can possibly be derived from the presented formalism, by approximating the memory term with higher order. Finally, the moment closure problem is not limited to radiative transfer. An important question is to which extent the ideas presented here can be applied to the Boltzmann equation of kinetic gas dynamics.

%=============================================================================================
\bibliographystyle{amsplain}
\bibliography{references_complete}
%=============================================================================================
\end{document}